\documentclass[12pt]{article}
\pdfoutput=1
\usepackage{putex}
\usepackage{graphicx}
\usepackage{epstopdf}
\usepackage{enumerate}
\usepackage{cite}
\usepackage{tensor}
\usepackage{slashed}
\usepackage{feynmf}
\usepackage{hyperref}
\usepackage{float}
\usepackage[bottom]{footmisc}
\usepackage[utf8]{inputenc}
\usepackage[margin=2.3cm]{geometry}

\usepackage[usenames,dvipsnames,svgnames,table]{xcolor}

\hypersetup{
	pdfstartview={FitH},    
	pdftitle={Semi-classical  Virasoro blocks: proof of exponentation},    
	pdfauthor={Besken, Datta, Kraus},     
	colorlinks=true,       
	linkcolor=blue,          
	citecolor=blue!59!black! ,        
	filecolor=magenta,      
	urlcolor=blue!75!black,           
	linktoc=all
}

\makeatletter
\g@addto@macro\bfseries{\boldmath}
\makeatother

\numberwithin{equation}{section}

\def\s{\sigma}

\def\a{\alpha}

\def\wb{\overline{w}}

\def\Lc{{\cal L}}

\newcommand {\be} {\begin {equation}}
\newcommand {\ee} {\end {equation}}
\newcommand {\nn} {\nonumber}
\newcommand {\bes} {\begin {equation*}}
\newcommand {\ees} {\end {equation*}}

\newcommand{\beq}{\begin{equation}}
\newcommand{\eeq}{\end{equation}}

\newcommand{\cS}{\mathcal{S}}

\newcommand{\cV}{\mathcal{V}}

\newcommand{\p}{\partial}

\def\be{ \begin{equation} }
\def\ee{ \end{equation} }

\def\b{\beta}
\def\l{\lambda}

\def\m{\mu}
\def\wb{\overline{w}}

\def\tf{\tilde{\phi}}
\def\tA{\tilde{A}}

\def\tS{\tilde{S}}

\newcommand{\bea}{\begin{eqnarray}}
\newcommand{\eea}{\end{eqnarray}}

\newcommand{\no}{\nonumber}

\def\wb{\overline{w}}

\def\ub{\overline{u}}

\def\rt{\rightarrow}

\def\pd{\partial}

\renewcommand{\>}{\rangle}

\begin{document}
	
	\institution{UCLA}{ \quad\quad\quad\quad\quad\quad\quad\ ~ \, $^{1}$Mani L. Bhaumik Institute for Theoretical Physics
		\cr Department of Physics \& Astronomy,\,University of California,\,Los Angeles,\,CA\,90095,\,USA}

	\institution{UvA}{ \quad\quad\quad\quad\quad\quad\quad\quad\quad\quad\quad\quad \,
	$^{2}$Institute for Theoretical Physics, \cr
	\quad University of Amsterdam, Science Park 904, 1098 XH Amsterdam, The Netherlands}

	\title{Semi-classical  Virasoro blocks: \\  proof of exponentiation
	}
	
	\authors{Mert Be\c sken$^{1,2}$, Shouvik Datta$^{1}$ and Per Kraus$^{1}$  \\ \vskip0.4cm \fontfamily{lmtt}\fontsize{10}{12}\selectfont \textbf{m.besken@uva.nl,~~shouvik@ucla.edu,~~pkraus@ucla.edu} \\ \vskip0.4cm}
	
	\abstract{Virasoro conformal blocks are expected to exponentiate in the limit of large central charge $c$ and large operator dimensions $h_i$,
 with the ratios $h_i/c$ held fixed. We prove this by employing the oscillator formulation of the Virasoro algebra and its representations. The techniques developed are then used to provide new derivations of some standard results on conformal blocks.
 }
	
	\date{}
	
	\maketitle
	\setcounter{tocdepth}{2}
	\begingroup
	\hypersetup{linkcolor=black}
	\tableofcontents
	\endgroup
	

\section{Introduction}

Conformal blocks in 2d CFTs are fixed by Virasoro symmetry. However, closed form expressions are known only in some special cases. A general feature of the semi-classical limit,
\begin{align}
c \to \infty, \quad h_i,h \to \infty, \quad \frac{h_i}{c}, \frac{h}{c} \ \text{~fixed}~,
\end{align}
is that the conformal block is believed to  exponentiate, i.e.~it takes the form \cite{Zamolodchikov:1985ie}
\begin{align}\label{eq:expo}
\cV(h_i,h,c;z) \approx \exp\left[-\frac{c}{6}f\left(\frac{h_i}{c}, \frac{h}{c};z\right)\right]~.
\end{align}
Here $c$ is the central charge, $h_i$ are conformal dimensions of the external operators, $h$ is the conformal dimension of the exchanged primary and $z$ is the cross-ratio. Although there is compelling evidence for \eqref{eq:expo}, a first principles derivation of this well-known formula is lacking. The aim of this paper is to close this gap.

An intuitively appealing, but somewhat heuristic,  argument for exponentiation is provided by Liouville theory.  At large $c$, correlation functions of heavy primary operators may be computed using the saddle point approximation to the Liouville path integral.  Assuming that the saddle point picks out a particular Virasoro block, together with the large $c$ behavior of  the DOZZ structure constants in this regime \cite{Zamolodchikov:1995aa}, the result follows.
A strong check of \eqref{eq:expo} comes from evaluating the conformal block in a power-series expansion in $z$ to high orders using Zamolodchikov's recursion relation \cite{Zamolodchikov:1985ie}. Further indications of this exponential feature also arise from the AGT correspondence  \cite{Alday:2009aq}, in which the coefficients of the cross-ratio expansion are combinatorially related to the instanton part of the Nekrasov partition function \cite{Nekrasov:2013xda,Alba:2010qc}.
This exponentiation property has found many applications in recent times --- e.g. for deriving heavy-light conformal blocks using the monodromy method \cite{Fitzpatrick:2014vua}, and in various applications in the context of AdS$_3$/CFT$_2$ which relate to chaos, entanglement and thermalization \cite{Hartman:2013mia,Roberts:2014ifa,Asplund:2014coa,Fitzpatrick:2015zha}.

In this paper, we make use of the oscillator representation of the Virasoro algebra to prove \eqref{eq:expo}.  Our proof is direct and explicit.  The oscillator approach to calculating conformal blocks was developed in \cite{zamolodchikov1986two}, where it was used to derive a closed form expression for the block $\cV(\frac{1}{16},h_p,1;z)$. In the oscillator formalism, the Virasoro generators are represented as differential operators acting on an infinite collection of complex variables. CFT states are represented as holomorphic ``wavefunctions" of these variables.  A Virasoro block is expressed as the inner product between two wavefunctions, each representing a state  $O_i O_j |0\rangle$ projected onto a single representation $h$ corresponding to the exchanged operator.  Our derivation of \eqref{eq:expo}   will proceed by showing that the wavefunctions exponentiate in the semi-classical limit. The inner product, which yields the block, then boils down to evaluating  integrals via the saddle-point approximation.

The outline of this paper is as follows. In Section \ref{sec:osc} we provide a short review of the oscillator formalism and the procedure for obtaining conformal blocks.  Section \ref{sec:proof} contains the proof of the exponentiation of the block in the semiclassical regime. We utilize the oscillator machinery to work out few examples of conformal blocks in Section \ref{sec:eg}. Appendix \ref{app:} has some technical details on the proof of exponentiation.


\section{Virasoro blocks from the oscillator formalism}
\label{sec:osc}
In this section we briefly review the the oscillator representation of the Virasoro algebra \cite{zamolodchikov1986two} and its application to the computation of conformal blocks. A detailed discussion of this formalism and its applications, along with its derivation from the linear dilaton theory can be found in Appendix A of \cite{Besken:2019bsu}.

\subsection{Oscillator formalism}
The Hilbert space of a 2d CFT is organized in representations of two copies of the Virasoro algebra formed by modes of the stress tensor which obey
\bea
[L_m,L_n]= (m-n)L_{m+n} +{c\over 12}(m^3-m)\delta_{m,-n}~.
\eea
We focus on the holomorphic sector. The  Virasoro module associated with the primary state $|h\>$ (where $h$ denotes the $L_0$ eigenvalue) is generated by the action of the raising operators $L_{n<0}$.  A generic state $|f\>$ in this module is described by the wavefunction $f(u) \equiv \<u | f \>$ where $u$ denotes the infinite collection of oscillator variables $\{u_1, u_2, \ldots\}.$ Each $u_i$ is a complex coordinate and $f(u)$ are holomorphic functions on $\mathbb{C}^{\infty}$. The action of the Virasoro generators on  wavefunctions is given by $\<u|L_k|f\> = l_k f(u)$, where
\begin{equation}
\label{viro}
\begin{aligned}
l_0 & = h+ \sum_{n=1}^\infty n u_n {\p\over\p u_n}, \cr
l_k &= \sum_{n=1}^\infty n u_n {\p\over\p u_{n+k}} -{1\over 4} \sum_{n=1}^{k-1} {\p^2 \over \p u_n \p u_{k-n} }+(\mu k+i\lambda){\p \over \p u_k}~,\quad\quad k>0 \cr
l_{-k} &= \sum_{n=1}^\infty (n+k) u_{n+k} {\p\over\p u_{n}} - \sum_{n=1}^{k-1}n(k-n) u_n u_{k-n}+2k(\mu k-i\lambda)u_k~,\quad\quad k>0~.
\end{aligned}
\end{equation}
where the real parameters $\l$ and $\m$ are related to the central charge and conformal dimension of the primary
\bea
\label{cmhl}
c= 1+24 \mu^2,~~h=\lambda^2+\mu^2~.
\eea
A state in the dual representation is described by the wavefunction $\overline{f(u)} \equiv \<f|\ub\>$ where bar acts on $\m$ and $\l$ trivially, sends $i \to -i,$ and sends oscillators $u_n$ to their antiholomorphic counterparts $\ub_n.$ This translates to the action of $L_n$ as
\bea \label{oust}\langle f |L_n |\ub\rangle = \overline{\langle u|L_{-n}|f\rangle} = \overline{ l_{-n} f(u)} = \overline{l}_{-n}\overline{f(u)}~.
\eea
The inner product
\bea
\label{innp}
\big(f(u),~g(u)\big) =   \int\! [du] \overline{f(u)}g(u)~,~~~[du] =  \prod_{n=1}^\infty d^2u_n {2n \over \pi} e^{-2n  u_n\ub_n}~,
\eea
realizes the adjoint relations, $l_n^\dagger = \overline{l}_{-n}$, resulting in a unitary representation.  The integration measure is normalized so that $(1,1)=1$.

A generic descendant state at level $N=\sum_j j m_j$ is a sum of monomials $u_1^{m_1} u_2^{m_2} \ldots u_m^{m_N}.$ This follows simply from the definition of $l_0$ in \eqref{viro}. The inner product of a monomial $u_1^{m_1} u_2^{m_2} \ldots$ with itself is built out of
\begin{align}\label{snor}
S_{j,k}={2j\over \pi}\int_{\mathbb{C}} du_jd\ub_j ~ e^{-2ju_j\ub_j}|u_j|^{2k}=  (2j)^{-k}\Gamma(k+1)~,
\end{align}
and the inner product between any two distinct monomials is zero; these monomials thereby form an orthogonal basis.

\subsection{States and wavefunctions}
Our focus will be on wavefunctions describing states created by primary operators acting on the vacuum, $O_{h_1}(z_1) \ldots O_{h_n}(z_n)|0\rangle$.  In \cite{Besken:2019bsu}, the 1-point wavefunction for the primary $O_h(z)$ was computed
%
\bea
\langle u|O_h(z)|0\rangle = \exp \Big\{{2(\mu-i\lambda) \sum_{n=1}^\infty z^n u_n}\Big\}~.
\eea
Similarly, we define 2-point wavefunctions\footnote{As we have seen, a set of oscillators $u$ comes with a label $h$ referring to the highest weight of their associated representation. This label appears as the subscript in the 2-point wavefunction, and implies a projection onto the associated conformal family. That is, $ \psi_h(z_1,z_2,U) $ represents the state $P_{h}O_{h_1}(z_1) O_{h_2}(z_2)|0\rangle$, where $P_h$ is the projection operator onto the representation labelled by $h$. }
\bea\label{psi-def} \psi_h(z_1,z_2,u) = \langle u|O_{h_1}(z_1) O_{h_2}(z_2)|0\rangle~,\quad
\chi_{h}(z_3,z_4,\ub) = \langle 0 |O_{h_3}(z_3)O_{h_4}(z_4)|\ub\rangle~.
\eea
To compute $\psi_h(z_1,z_2,u)$ we use the simple relation
\bea \label{diff}
0=\<u|L_k O_{h_1}(z_1)O_{h_2}(z_2) - [L_k, O_{h_1}(z_1)O_{h_2}(z_2)] |0\>,~~~k \geq -1~.
\eea
The Virasoro generators act on primary operators on the plane as
\bea
[L_k, O_h(z)] = -\Lc_k O_h(z)~,~~~\Lc_k=-z^{k+1}\p_z-(k+1)hz^k~,
\eea
and from \eqref{diff} and \eqref{psi-def} this implies
\bea
\label{inst}
\left(l_k + \Lc_k^{(1)} + \Lc_k^{(2)}\right) \psi_h(z_1,z_2,u) = 0,~~~k \geq -1~.
\eea
Conjugating using (\ref{oust}), this implies for $\chi_h(z_3,z_4,\ub)$
\bea
\left( \overline{l}_{-k} - \Lc_k^{(3)} - \Lc_k^{(4)}\right) \chi_h(z_3,z_4,\ub) = 0,~~~k \geq -1~.
\eea\vspace{-1.4cm}
\subsection{Virasoro blocks}
To compute Virasoro blocks we take the  inner product \eqref{innp} of the wavefunctions $\psi_h$ and $\chi_h$. Denoting the cross-ratio by $z$, a generic Virasoro block is given by
\begin{equation}
\label{vibl}
\begin{aligned}
\cV(c,h,h_i;z) = z^{h-h_1-h_2} V(c,h,h_i;z) &= \langle 0| O_{h_4}(\infty)O_{h_3}(1)P_{h} O_{h_1}(z)O_{h_2}(0) |0\rangle \\  &=  \int [du]\, \chi_{h}(1,\infty,\ub)\, \psi_{h}(z,0,u)~.~\cr
\end{aligned}
\end{equation}
We focus on $\psi_h(z,0,u)$ since $\chi_{h}(1,\infty,\ub) = \overline{\psi_{h}(1,0,u)} |_{h_{1,2}\to h_{3,4}}.$ The $k=0$ equation of (\ref{inst}) fixes the $z$-dependence
\bea
\psi_{h}(z,0,u) = z^{h-h_1-h_2}F(\eta)
\eea
where $\eta$ denotes the collection of rescaled oscillator variables, \footnote{We use this notation throughout.  Omission of a subscript is meant to denote the infinite collection of these oscillators.}
\bea
\label{etas}
\eta_m = z^m u_m,~~m=1,2,\cdots~.
\eea
If desired, the dependence on the second $z$-argument can be restored using $\psi_{h}(z_1,z_2,u) = z_{12}^{h-h_1-h_2}F(\eta, {z_2\over z_1})$, which follows from the $k=-1$ equation.\footnote{We use the shorthand $F(\eta) \equiv F(\eta,0).$}

 For $k \geq 1$ we have
\bea
\label{feqn}
\left[\sum_{n=1}^\infty n\eta_n \left({\p \over \p \eta_{n+k}} - {\p \over \p \eta_{n}}\right) - {1\over 4} \sum_{n=1}^{k-1} {\p^2\over \p \eta_n \eta_{k-n}} + (\mu k +i \l) {\p \over \p \eta_{k}} +h_2-kh_1-h \right] F(\eta) = 0~.\cr
\eea
$F(\eta)$ has a natural decomposition in terms of the descendant levels of monomials in the variables $\eta_m$. As we shall see shortly, this corresponds to the small cross-ratio expansion for the block. Specifically
\begin{align}
\label{feta}
F(\eta) &= \sum_{m=0}^\infty F_m  
= 1 + \underbrace{\left( \phi_{\{1^1\}} \eta_1  \right)}_{F_1} + \underbrace{\left( \phi_{\{1^2\}} \eta_1^2 + \phi_{\{2^1\}} \eta_2 \right)}_{F_2} + \underbrace{\left( \phi_{\{1^3\}} \eta_1^3 + \phi_{\{1^1,2^1\}} \eta_1 \eta_2 + \phi_{\{3^1\}} \eta_3 \right)}_{F_3} + \cdots~.
\end{align}
Here we denote a partition of an integer $m$ by a set of pairs of integers $\{j^{k_j}\};$  $m=\sum_j jk_j.$ For example, a specific partition of $5$ is $\{1^3,2^1\}.$

The wavefunction can be computed
 by plugging  (\ref{feta}) into the differential equation (\ref{feqn}), collecting coefficients of monomials $\eta_1^{m_1} \eta_2^{m_2} \ldots$,  and then  setting each coefficient to zero. This gives a linear system of equations for $\phi_{\lbrace j^{k_j}\rbrace}$'s at each level. For example, at level 1 we find
\bea\label{one-one}
\phi_{\{1^1\}} = {h+ h_1 - h_2 \over \m+i\l}~.
\eea
Determining $F(\eta)$  level-by-level and taking the inner product (\ref{vibl}), we arrive at the cross-ratio expansion of the  Virasoro block,
\begin{align}
V(c,h,h_i;z) &= \sum_{n=0}^\infty V_n z^n = \sum_{n=0}^\infty z^n \sum_{\{j^{k_j}\}} \phi_{\{j^{k_j}\}} \tf_{\{j^{k_j}\}} \prod_j S_{j,k_j}~,\no \\
&= 1 + \phi_{\{1^1\}} \tf_{\{1^1\}} S_{1,1}  z + \left( \phi_{\{1^2\}} \tf_{\{1^2\}} S_{1,2} + \phi_{\{2^1\}} \tf_{\{2^1\}} S_{2,1}  \right) z^2 + \cdots .
\end{align}
Here tilde denotes the combination of operations $i \to -i,~h_{1,2} \to h_{3,4}$ and $S_{j,k_j}$ is given by (\ref{snor}).   The steps described above are simple to implement on a computer.

The standard method to obtain the  conformal block in a cross-ratio expansion is to solve the Zamolodchikov recursion relations \cite{Zamolodchikov:1985ie}.    For obtaining high order numerical results, this approach is much more efficient than using wavefunctions.  On the other hand, the derivation of the recursion relation is not entirely rigorous (e.g.~see the discussion in \cite{Perlmutter:2015iya}), while the wavefunction based derivation is completely transparent.  Verifying agreement between the two approaches, as we have done, is reassuring.

\section{Exponentiation of semi-classical  Virasoro blocks}
\label{sec:proof}

In this section we prove the exponentiation \eqref{eq:expo} of  Virasoro blocks in the limit $c\to\infty$ with the ratios of operator dimensions, $h_i/c$ and $h/c$, held fixed.


Our starting point is (\ref{feqn}). The $c\to\infty$ limit can be implemented by taking $\mu\to\infty$, for which we have $c\approx 24\mu^2$.  We perform another rescaling of the oscillators,
\bea
\eta_m = \m \s_m~,
\eea
and define
\bea
\l = \a \m,~~h_{1,2}= \m^2 g_{1,2}~.
\eea
Here, $\alpha$ and $g_i$ are parameters that are held fixed in the  large $\mu$ limit.
The conformal dimension of the exchanged primary is then, $h=\mu^2+\lambda^2=(1+\alpha^2)\mu^2$. The equations \eqref{feqn} now take the form
\begin{align}\label{fs-eqn}
\left(\sum_{n=1}^\infty n \s_n \left({\p \over \p \s_{n+k}} - {\p \over \p \s_{n}}\right) - {1\over 4\m^2} \sum_{n=1}^{k-1} {\p^2\over \p \s_n \s_{k-n}} + \left(k  +i\a\right) {\p \over \p \s_{k}} + \m^2\gamma_k \right) F(\s) = 0
\end{align}
where we have defined $\gamma_k=\left( g_2-kg_1-1 -\a^2 \right)$.
Our goal is to show that for large $\mu$  this system of equations admits a solution, to all orders in the cross-ratio expansion, of the form 
\bea\label{ansatz}
F(\s) = e^{\m^2 S(\s)}~
\eea
where we suppressed dependence on $g_{1,2}$ and $\a$ for brevity. Plugging \eqref{ansatz} into \eqref{fs-eqn} and keeping the  leading terms in the large $\m$ limit, we get\footnote{We note that subleading corrections in the $1/\mu^2$ expansion can be systematically calculated order-by-order using this procedure.}
\begin{align}\label{Seqn}
	\sum_{n=1}^\infty n \s_n \left({\p S \over \p \s_{n+k}} - {\p S \over \p \s_{n}}\right) - {1\over 4}\sum_{n=1}^{k-1}{\p S \over \p \s_{n}} {\p S \over \p \s_{k-n}}+(k+i \alpha) {\p S \over \p \s_{k}}+\gamma_k=0~.
\end{align}
The above differential equation is now first order in oscillator derivatives. 
As for $F(\eta)$,  $S(\s)$  may be expanded in the oscillators,
\bea
\label{sesi}
\hspace{-.3cm}
S(\s) = \sum_{m=1}^\infty A_m \s_m + \sum_{m,n=1}^\infty B_{m,n} \s_m \s_n + \sum_{m,n,l=1}^\infty C_{m,n,l} \s_m \s_n \s_l + \cdots~,
\eea
where the coefficients $A_m,\,B_{m,n},\,C_{m,n,l},\, \cdots$ depend on $g_{1,2},\a.$ In Appendix \ref{app:} we detail the procedure for computing these, and prove that a  solution exists for generic values of the parameters.\footnote{This statement is not obvious, since the system of equations resulting from plugging (\ref{sesi}) into (\ref{Seqn}) naively appears to be overconstrained.  However, reorganizing the expansion establishes the existence of a solution. See Appendix \ref{app:} for further discussion.} This establishes  that the wavefunction takes the form (\ref{ansatz}) in the large $\mu$ limit.

By using the inner product (\ref{vibl}) and reverting to the original oscillator variables $u$, the corresponding Virasoro block takes the form
\begin{align}
V(\a, g_i;z) = \int \left( \prod_{n=1}^\infty d^2u_n \frac{2n}{\pi}  \right) \exp\left\{ -2 \sum_m m  u_m\ub_m  +\m^2 \left(  \sum_{m=1}^\infty A_m z^m {u_m\over \m}  +  \sum_{m=1}^\infty \tA_m {\ub_m\over \m} + \ldots\right) \right\},
\end{align}
where we recall tilde indicates the replacements $i \to -i,~g_{1,2} \to g_{3,4}.$ The ellipsis denotes terms of higher order in ${u_m\over \m}$ and ${\ub_m\over \m}.$ This suggests the change of variables $u_m, \ub_m \to \m u_m, \m \ub_m$ and we can write the Virasoro block in the following form \footnote{We discard $z$-independent factors; the overall normalization of the conformal block can be fixed by the OPE limit, $\cV(z\to 0)\approx z^{h-h_1-h_2}$.}
\begin{align}
\label{V}
&V(\a, g_i;z) \sim \int \left( \prod_{n=1}^\infty d^2u_n \frac{2n}{\pi} \right) e^{\m^2 {\cal I}(\alpha,g_i;u,\ub,z) }, \quad \nn \\
&{\cal I}(\alpha,g_i;u,\ub,z)= -2 \sum_m m  u_m\ub_m  +  S(\eta) + \tS(\ub) ~.
\end{align}
For large $\m$  the integral is dominated by the saddle point at $(u_m=w_m, \overline{u}_m=\overline{w}_m)$, with
\begin{align}\label{saddle-point}
\wb_m = {1\over 2m} \bigg.{\p S(\eta)\over \p u_m}\bigg|_{u=w},~~~w_m = {1\over 2m} \bigg.{\p \tS(\ub)\over \p \ub_m}\bigg|_{\ub=\wb}~.
\end{align}
Plugging these values back  in the integrand of \eqref{V} and denoting the ``on-shell action"  as $\cS(h_i/c,h/c;z)=-\frac{1}{4}{\cal I}(\alpha,g_i;w,\overline w,z)$, we have
\begin{align}
V(\a, g_i;z) 
\approx \exp \left\{ -\,\frac{c}{6}\,{\cal S}\left(\frac{h_i}{c},\frac{h}{c};z\right)\right\}~.
\end{align}
The power-law prefactor,  $z^{h-h_1-h_2}$, in \eqref{vibl} takes this same exponential form.  This completes our proof that the Virasoro block exponentiates in the semi-classical limit.

Stepping back, the key point is that the defining equations for the wavefunction \eqref{feqn}  admit an exponential ansatz in the semi-classical regime. The equations then reduce to a linear system  governing the coefficients of the function $S(\sigma)$.  The remaining step is to prove existence and uniqueness of a solution to this linear system of equations.  This is shown in Appendix \ref{app:} by rearranging the equations.  We have verified that the function ${\cal S}\left(\frac{h_i}{c},\frac{h}{c};z\right)$ agrees with corresponding expressions obtained from the recursion relation and the monodromy method.  The monodromy method is an efficient tool for computation but relies on the assumption of exponentiation;  our proof removes this assumption and puts this on firm footing.

\def\bS{\bar{S}}
\def\bs{\bar{\s}}
\def\ausricht{\begin{aligned}}
\def\endeausricht{\end{aligned}}

\section{Examples}
\label{sec:eg}
We now work out some concrete examples to illustrate the procedure to compute Virasoro blocks using the oscillator formalism. The results about to be derived are not new, but they serve as practical demonstrations of this approach.

\subsection{Perturbatively heavy vacuum block}

Our first example is the vacuum block ($h=0$) in the limit  $c\rt \infty$ with $h_i/c$ fixed. This limit requires considering imaginary values of $\l$ (\ref{cmhl}), but this would spoil the adjoint relations $l_n^\dagger = \overline{l}_{-n}$ and result in a non-unitary representation. The correct procedure is to perform analytic continuation in $h$ as discussed in Appendix A.3 of \cite{Besken:2019bsu} which we summarize here. The key point is to observe that $i$ appears only in the combination $i\l$ in (\ref{feqn}) which fix the wavefunction $\psi_h.$ Replacing the rule $i \to -i$ for obtaining the conjugate wavefunction with $\l \to -\l$  we can consider imaginary values of $\l.$ More precisely, given the two possible imaginary values $\l^{\pm} = \pm i\sqrt{\m^2-h}$ of $\l$ for $h<\m^2$ we plug  $\l^+$ into (\ref{feqn}) and denote the solution $\psi^+_h.$ We pair this up with $\psi^-_h$ which is the solution of the same equations with replacements $h_{1,2} \to h_{3,4},~u \to \ub,$ and $\l^+ \to \l^-,$ which results in the block
\begin{align}
\cV(c,h,h_i;z) =  \int [du]\, \psi^-_{h}(1,\infty,\ub)\, \psi^+_{h}(z,0,u).
\end{align}
Since Virasoro blocks are rational functions of $h$ in a cross ratio expansion, we are free to exchange $+ \leftrightarrow -$ in the above formula.

Returning to our example for the vacuum block we have $h_1=h_2$ and $h_3=h_4$.   Alternatively, one can access this regime by expanding ${\cal S}$ to lowest order in $h_{1,3}$.

Since we are considering vacuum exchange, we choose $\alpha=i$ for the wavefunction $e^{\mu^2 S^+(g_1;\s)}$ and $\alpha=-i$ for its conjugate $e^{\mu^2 S^-(g_3;\bs)}$. The differential equations \eqref{fs-eqn} read
\begin{align}\label{LLdiff}
	\ausricht
	\sum_{n=1}^\infty n\sigma_n \left(\frac{\pd S^+}{\pd \sigma_{n+k}}-\frac{\pd S^+}{\pd \sigma_{n}}\right)- \frac{1}{4}\sum_{n=1}^{k-1} \frac{\pd S^+}{\pd \sigma_n}\frac{\pd S^+}{\pd \sigma_{k-n}} + (k-1) \frac{\pd S^+}{\pd \sigma_k} -(k-1)g_1&=0~,\\
	\sum_{n=1}^\infty n\bs_n \left(\frac{\pd S^-}{\pd \bs_{n+k}}-\frac{\pd S^-}{\pd \bs_{n}}\right)- \frac{1}{4}\sum_{n=1}^{k-1} \frac{\pd S^-}{\pd \bs_n}\frac{\pd S^-}{\pd \bs_{k-n}} + (k+1) \frac{\pd S^-}{\pd \bs_k} -(k-1)g_3&=0~.
	\endeausricht
\end{align}
We desire to obtain $\log \cV$ to  linear order in each of the conformal dimensions $g_{1,3}$, as higher order terms will be accompanied by inverse powers of $\mu$ and hence vanish in the limit under consideration.  Therefore, we look for solutions of the form $S^+(g_1;\sigma)= g_1 s^+(\sigma)+O(g_1^2)$ and  $S^-(g_3;\sigma)= g_3 s^-(\sigma)+O(g_3^2)$.\footnote{There is  an interesting subtlety associated with this assumption, which is discussed at the end of this section.}
It is straightforward to see from the saddle-point equations \eqref{saddle-point} that the saddle point values $(w_m,\overline{w}_m)$ vanish linearly with $g_{1,3}$ as $g_{1,3}\rt 0$.    Thus, terms in $S^+$ and $S^-$ higher than linear order in the oscillators make no contribution to the conformal block at linear order in $g_{1,3}$.  Hence we need only determine the parts of  $S^+$ and $S^-$ linear in oscillators.  These are easy to find, simply by using the last two terms in the equations (\ref{LLdiff}), and we obtain\footnote{The $\sigma_1$ term in $S$ actually has an undetermined coefficient, and we have simply chosen a specific value.  The choice is immaterial, since  $\sigma_1$ will vanish at the saddle point.}
\begin{align}
\label{Slin}
S^+\approx g_1\sum_{n=1}^{\infty} \s_n ~,\qquad  S^- \approx g_3\sum_{n=1}^{\infty} \frac{n-1}{n+1}\bs_n ~.
\end{align}
We now rescale the oscillator variables as in the previous section and plug this back into \eqref{V}. The ``action" reads
\begin{align}
{\cal I}
\approx & -2 \sum_m m  u_m\ub_m  +   {g_1\over \mu}\sum_{n=1}^{\infty} u_n z^n   + {g_3\over \mu}\sum_{n=1}^{\infty} \frac{n-1}{n+1}\ub_n ~.
\end{align}
Upon extremization, we have
\begin{align}
\label{Vph}
V(\a, g_i;z) &\approx \exp\left\{ \mu^2 {\cal I} \right\}= \exp \bigg\lbrace {\mu^2 g_1 g_3 \over 2}\sum_{n=1}^\infty \frac{n-1}{n(n+1)}z^n  \bigg\rbrace = \exp \bigg\lbrace {2h_1 h_3 \over c}z^2 {}_2 F_1(2,2,4,z)  \bigg\rbrace~.
\end{align}
The term in the exponent is the global block for stress tensor exchange.  This result was derived in \cite{Fitzpatrick:2014vua} by the monodromy method and also by a direct summation over descendants of the vacuum.

\subsubsection*{Solution branches}
\label{subtle}

We now comment on a subtlety that arises upon solving the first equation in (\ref{LLdiff}) for $S^+$. As always, we are interested in solutions taking the form of a power series expansion in oscillators, where the constant term can be chosen to vanish.  The subtlety has to do with the observation that there are two distinct branches of such solutions.   Suppose we write the first few terms in the level expansion as $S^+= C_1 \sigma_1 + C_2 \sigma_2 +\cdots $.  Plugging in, we find $C_1=C_2 = 2(1\pm \sqrt{1-g_1})$.  The $\pm =-$ branch\footnote{Hopefully this does not cause confusion with the $\pm$ in $S^\pm$!} corresponds to the solution (\ref{Slin}).   What is perhaps surprising is that, as is easily checked, it is the other branch  that is obtained by solving for the first few terms in the wavefunction with general parameters (e.g.~equation~\eqref{one-one}), and then taking the semiclassical limit at the end.  So it may appear that we have chosen the ``wrong" or ``disconnected'' branch.   However, this turns out not to matter.   The reason is that the two solutions are related to each other by analytically continuing $g_1$ around the branch point at $g_1=1$.   But as mentioned Virasoro blocks are known to be rational functions of the conformal dimensions at each order in the cross-ratio expansion.  Hence branch cuts are absent in the Virasoro block, and so the two wavefunctions will yield the same result for the block.  The reason for our choice of branch is that the linearly vanishing small $g_1$ behavior allowed us to restrict to terms linear in oscillators, while this simplification would not be present for the other branch.

\subsection{Heavy-light block}

In our next example, we generalize (\ref{Vph}) by computing ${\cal I}$ to first order in $g_1$ but to all orders in $g_3$.  The Virasoro block in this regime was computed in \cite{Fitzpatrick:2014vua}, and here we show how this result emerges straightforwardly in our approach.

We consider vacuum exchange.
The $g_1$ part of the computation is unchanged, so we have $S^+=g_1\sum_{n=1}^\infty \sigma_n$.   From (\ref{V}) the action to be extremized has the form
\bea
{\cal I}= -2 \sum_m m  \sigma_m \bs_m  + g_1 \sum_{n=1}^\infty \sigma_n + S^-(g_3;\bs)~.
\eea
The saddle point equations give
\bea
\bs_m={g_1\over 2m}~,\quad \sigma_m ={1\over 2m} {\p S^- \over\p \bs_m}\Big|_{\bs_m=g_1/2m}~.
\eea
We see that only the part of $S^-$ linear in oscillators will contribute to ${\cal I}$  at linear order in $g_1$.   We denote
\bea
S^-_{\rm lin}(g_3,\bs_m)= \sum_{m=2}^\infty A_m(g_3) \bs_m
\eea
%
At the saddle point the action has the form (after restoring the cross-ratio dependence)
\bea\label{hl-action}
{\cal I} = g_1 \sum_{m=2}^\infty {A_m  \over 2m}z^m~.
\eea
We now turn to the computation of $A_m(g_3)$. The second equation of (\ref{LLdiff}) implies the following recursion relation for the coefficients $A_m$,
\bea
-{1\over 4} \sum_{n=1}^{k-1} A_n A_{k-n}+(k+1)A_k-(k-1)g_3=0~.
\eea
Defining the generating function $A(x) = \sum_{n=2}^\infty A_n x^n$ this becomes the differential equation
\bea
{d\over dx}\big(xA (x)\big)={1\over 4} A(x)^2 +{g_3x^2 \over (1-x)^2}~.
\eea
This is a special case of the Riccati equation and the solution with small $x$ behavior $A \sim x^2$ is
\bea
A(x) = {2\over 1-x} \left[ 2-x + \alpha_3 x\left( 1-{2\over 1-(1-x)^{\alpha_3}}\right)\right]~,\quad \alpha_3=\sqrt{g_3-1}~.
\eea
We then compute ${\cal I}$, equation \eqref{hl-action}, via
\bea
{\cal I} = {g_1\over 2} \int {A(z) \over z }dz~.
\eea
This is readily evaluated (and the integration constant is fixed by demanding $\cV(z\to 0)\approx z^{-2h_1}$), yielding the Virasoro vacuum block
\bea
{\cal V}(z) \approx z^{-2h_1} V(z) \approx \big(1-w(z)\big)^{h_1(1-{1\over \alpha_3}) } \left({w(z)\over {\alpha_3}}\right)^{-2h_1}~,\quad  w(z)= 1-(1-z)^{\alpha_3}~,
\eea
in agreement with the result obtained in \cite{Fitzpatrick:2014vua}.

\subsection{Blocks with heavy exchange}
 For our final example, we consider Virasoro blocks with heavy intermediate exchange but at \textit{finite} central charge.  That is, we take the exchanged dimension $h\rt \infty$ but with the external dimensions $h_i$ and central charge $c$ held fixed.  In our notation this corresponds to taking $\lambda \rt \infty$ with other parameters held fixed.
The behavior of the  block in this regime is given by the well-known formula $\cV(q)\approx(16q)^h$, where $q=q(z)$ is defined below.  To our knowledge, the only existing  derivation of this is at \textit{large} central charge using the monodromy method \cite{Harlow:2011ny}.  Although this is an example beyond the semi-classical regime, the structure of the equations for the wavefunctions will turn out to be very similar.

Our starting point is (\ref{feqn}). We rescale the oscillators as $\eta_m = \l \s_m$
\begin{align}
\left(\sum_{n=1}^\infty n \s_n \left({\p \over \p \s_{n+k}} - {\p \over \p \s_{n}}\right) - {1\over 4\l^2} \sum_{n=1}^{k-1} {\p^2\over \p \s_n \s_{k-n}} + \left({\m k\over \l}  +i\right) {\p \over \p \s_{k}} + \delta_k \right) F(\s) = 0~,
\end{align}
with $\delta_k=h_2-kh_1-\m^2 -\l^2$.
Next, we plug in the ansatz
\bea
F(\s) = e^{\l^2 S(\s)}~.
\eea
Taking the large $\l$ limit with everything else fixed and keeping leading terms, we have
\begin{align}
\label{sell}
\sum_{n=1}^\infty n \s_n \left({\p S \over \p \s_{n+k}} - {\p S \over \p \s_{n}}\right) - {1\over 4}\sum_{n=1}^{k-1}{\p S \over \p \s_{n}} {\p S \over \p \s_{k-n}}+i{\p S \over \p \s_{k}} -1=0~.
\end{align}
It can be seen that the analog of the expansion (\ref{sesi})  consistently trunctates at quadratic order in oscillators
\bea
\label{ssla}
S(\s) = \sum_{m=1}^\infty A_m \s_m + \sum_{m,n=1}^\infty B_{m,n} \s_m \s_n~.
\eea
Substituting this into (\ref{sell}), we arrive at the following equations for the coefficients $A_m$ and $B_{m,n}$
\begin{equation}
\begin{aligned}
&iA_k - {1\over 4}\sum_{n=1}^{k-1} A_n A_{k-n} -1 =0, \\
&2i B_{k,m} - \sum_{n=1}^{k-1} A_n B_{k-n,m} + m(A_{m+k}-A_m)=0,\\
&m(B_{l,m+k}-B_{l,m}) + l(B_{m,l+k}-B_{m,l}) - \sum_{n=1}^{k-1} \left( B_{m,n} B_{l,k-n} \right) =0~.
\end{aligned}
\end{equation}
An equivalent set of equations was solved in \cite{zamolodchikov1986two} while calculating the block $\cV(\frac{1}{16},h,1|z)$. The solution can be written in terms of the generating functions
\begin{align}
	A(p) = \sum_{m=1}^{\infty} A_m p^m,~~B(p,q)=\sum_{m,n=1}^{\infty}B_{m,n}p^m q^n,
\end{align}
as
\begin{equation}
	\begin{aligned}
	A(p) &= -2i\big[ (1-p)^{-{1\over 2}}-1 \big],\\
	B(p,q) &= -{1\over 2}pq \big[(1-p)(1-q)\big]^{-{1\over 2}} \big[ (1-p)^{1\over 2} + (1-q)^{1\over 2}\big]^{-2}~.
	\end{aligned}
\end{equation}
The remaining steps of the calculation are similar to those in \cite[Section 6]{zamolodchikov1986two} and we  borrow results from there in what follows.    After the rescaling, ${(u_m, \ub_m) \to (\l \sigma_m z^{-m}, \l \bs_m)}$, the Virasoro block in the large $\lambda$ limit is
\begin{align}\label{ZamoV}
{
V(\l ;z) \approx \int \left( \prod_{n=1}^\infty d^2\sigma_n \frac{2n}{\pi} \right) \exp\bigg\{   \l^2 \Big[ -2 \sum_m m  \sigma_m\bar\s_m z^{-m} + S(\sigma) + \tS(\bs) \Big] \bigg\}~,
}
\end{align}
where $S$ is read off from (\ref{ssla}) and the tilde now denotes $i \to -i$. Since the action is quadratic in the oscillators, this is a multi-dimensional Gaussian integral. Using $h\approx\lambda^2$ 
we obtain
\begin{align}
\label{VGH}
V(\lambda;z)\approx \exp\left[h\, H^T(1-2G)^{-1}H\right],
\end{align}
where we have ignored the determinant and Jacobian prefactors (as they are subleading contributions to $\log \cV$);  cross-ratio independent factors  are fixed by the OPE limit $z\to 0$ as before. The matrices $G$ and $H$ are
\begin{align}
H_j=z^{j/2}\frac{A_j}{\sqrt{2j}}~,\qquad G_{jk}=z^{j+k\over 2} \frac{B_{j,k}}{2\sqrt{jk}}~.
\end{align}
The object \eqref{VGH} was evaluated in \cite{zamolodchikov1986two} by mapping to a different problem with the same spectrum of eigenvalues. The result is
\begin{align}
V(z)\approx z^{-h}(16q)^h ~, \quad q=\exp\left[-\pi\frac{K(1-z)}{K(z)} \right]~,\quad K(z)={1\over 2}\int_0^1 {dt\over \sqrt{t(1-t)(1-zt)}} ~.
\end{align}
Combining with the power-law prefactor in \eqref{vibl} at large $h$, we get $\cV(q)\approx(16q)^h$. This is the expected result. As discussed above, we have now established the validity of this expression for arbitrary $c$.

\section{Discussion}

We proved that Virasoro blocks exponentiate in the semi-classical regime, and showed how to reproduce various results on Virasoro conformal blocks using the oscillator formalism. There are a number of directions  to extend our results which would be worthwhile to pursue.  For example, one could consider higher point conformal blocks, or blocks on higher genus surfaces and prove exponentiation formulas for these cases.  Other avenues include extensions to  super-Virasoro algebra and $\mathcal{W}$-algebras.

 Returning to the plane, further analysis of the equations governing the wavefunctions might lead to new analytical expressions for conformal blocks in certain parameter regimes.  For example, it would be valuable to have a closed form expression for the large $c$ wavefunction in the case of heavy external operators, $h_i\sim c$, and a light exchanged operator, $h\sim O(1)$.  Since expressions for  the Virasoro block in this regime are available \cite{Fitzpatrick:2015zha}, one suspects that the underlying wavefunctions can be determined.  Also of value would be further results on the wavefunctions and blocks with all operators heavy, and their connections to semiclassical gravity via the AdS$_3$/CFT$_2$ correspondence.

\section*{Acknowledgements}

P.K.~is supported in part by NSF grant PHY-1313986. M.B.~is supported by the Delta ITP consortium, a program of the 
Netherlands Organisation for Scientific Research (NWO) that is funded by 
the Dutch Ministry of Education, Culture and Science (OCW), and by the ERC Starting Grant GENGEOHOL.
	
\appendix	
\section{Consistency of the linear system of equations}
\label{app:}

In our approach, Virasoro blocks are obtained as the inner product of a pair of 2-point wavefunctions.   The coefficients in the oscillator expansion of these wavefunctions obey a system of linear equations.   In this appendix we show that this system of equations generically admits a unique solution, which may be obtained recursively order by order in the level expansion,  both for finite central charge and in the semi-classical limit.

We begin with the finite $c$ system of equations (\ref{feqn}). The function $F(\eta)$ has  $p(m)$ unknowns at each level $m$, $p(m)$ being the number of partitions of $m$.  We rearrange (\ref{feqn}) as\footnote{Here $l_k$ (\ref{viro}) are understood to be constructed out of $\eta_k$; the extra factor of $z^k$ compared to $u_k$ is inconsequential for the arguments that follow.}
\bea
l_k F = (l_0 +kh_1 -h_2) F~.
\eea
We plug in the level decomposition (\ref{feta}) and note the eigenvalue relation $l_0 F_m = (m+h) F_m$. This leads to
\begin{align}
	l_k F_m = \b_{k,m} F_{m-k},~~~\b_{k,m} = m-k +h +k h_1 -h_2~.
\end{align}
This equation can now be used recursively to determine the action of any string of $l_n$ generators on $F_m$.
Given a partition of $m =\sum_{i=1}^p j_i$, which we take to be ordered as $j_p \geq j_{p-1} \geq \ldots \geq j_1 \geq 1$, we have the following  action of a level-$m$  string of $l_{j_i}$ generators,
\bea
\label{pmeq}
l_{j_p} l_{j_{p-1}} \ldots l_{j_2} l_{j_1} F_m = \b_{j_{p},j_p} \b_{j_{p-1},j_p+j_{p-1}} \ldots \b_{j_2,m-j_1} \b_{j_1,m}~,
\eea
where we used $F_0=1$.
Clearly the number of such strings of $l_{j_i}$'s on the left hand side is $p(m)$, the same as the number of unknown coefficients in $F_m$, so we have the same number of equations as unknowns.    To establish the existence of a solution, we therefore need to show that the determinant of the corresponding $p(m)\times p(m)$ matrix is nonzero for generic parameters.
%
The  operators appearing on the left hand side of (\ref{pmeq}) are
\bea
\label{lops}
l_k = \sum_{n=1}^\infty n \eta_n {\p\over\p \eta_{n+k}} -{1\over 4} \sum_{n=1}^{k-1} {\p^2 \over \p \eta_n \p \eta_{k-n} }+(\mu k+i\lambda){\p \over \p \eta_k}~,\quad\quad k>0~.
\eea
Let us for  momentarily omit the first two sums and consider replacing $l_k$ by $\hat l_k =(\mu k+i\lambda){\p \over \p \eta_k}$.  The matrix corresponding to the system of equations (\ref{pmeq}) is then diagonal, with the diagonal entries being products of $ (\mu j_n +i\lambda)$ factors.   For   a representation with $h\geq 0$, the only possibility for  a vanishing factor occurs if $\lambda =i\mu$, which corresponds to the vacuum exchange, $h=0$.  This is related to the subtlety discussed in section \ref{subtle}.  In the following we will assume $h>0$; as far as the blocks are concerned there is no loss of generality since the $h\rt 0$ limit is smooth.   Therefore, if we replace $l_k\rt \hat{l}_k$ the matrix corresponding to the system of equations (\ref{pmeq}) has a non-zero determinant.  We now return to the original problem with $l_k$ operators.  Since $l_k-\hat{l}_k$ is independent of $\lambda$, the determinant is  a polynomial in $\lambda$ and its coefficient  of the term with the largest power of $\lambda$ is the same as in the $\hat{l}_k$ case.  Being a polynomial the determinant vanishes at a discrete set of $\lambda$ values (which may or may not be real and positive).  For generic values of $\lambda$ the determinant is non-vanishing and hence the system of linear equations generically admits a unique solution.


Next we turn to the semiclassical limit.  Without taking any limits, we first write $F=e^{\mu^2 S(\sigma)} $, $\eta_m= \mu \sigma_m$, with $S$ expanded in oscillator levels as $S=\sum_{k=1}^\infty S_k$, and plug into the system of equations (\ref{pmeq}).     $F_m$ takes the form $F_m = \mu^2 S_m + \ldots$, where the $\ldots$ involve terms built out of products of $S_{n<m}$.   At a given $m$, we regard (\ref{pmeq})  as a system  of equations determining $S_m$ in terms of the $S_{n<m}$, which may therefore be solved recursively.   As before, the number of equations matches the number of unknowns at each step of the recursion, so a unique solution will result provided the corresponding matrix determinant is nonvanishing.   This solution is of course the same as in our discussion above, just with a reorganized expansion in oscillators, hence uniqueness of the solution has already been proven.    Now, the semiclassical limit is obtained by making a large $\mu$  WKB-type replacement of the second derivative term in (\ref{lops}),
\bea
\ausricht
 {\p^2 \over \p \eta_n \p \eta_{k-n} } e^{\mu^2 S(\sigma) } &=  \mu^2 {\p S \over \p \sigma_n  } {\p S \over  \p \sigma_{k-n} }  + {\p^2 S \over \p \sigma_n \p \sigma_{k-n} } \\
 & \approx  \mu^2 {\p S \over \p \sigma_n  } {\p S \over  \p \sigma_{k-n} } ~.
 \endeausricht
\eea
Since the retained terms involve products of two $S_n$ factors, it is clear that the unknown  highest level term $S_m$ cannot appear.  Therefore, as far as the dependence on the highest level term $S_m$ in the semiclassical equations goes (which is what determines the matrix whose determinant we wish to study), we can omit the middle term in (\ref{lops}) and write
\bea
\label{lopsb}
l_k = \sum_{n=1}^\infty n \eta_n {\p\over\p \eta_{n+k}}+(\mu k+i\lambda){\p \over \p \eta_k}~,\quad\quad k>0~.
\eea
The argument that the corresponding matrix has non-vanishing determinant proceeds just like before: it is a polynomial in $\lambda$ and hence is generically non-zero.  This completes the proof of the existence and uniqueness of  wavefunctions in the semiclassical limit.   To gain further confidence in these arguments we have implemented the recursive solution of our equations on Mathematica, and a unique solution  indeed results order by order in the level expansion. The first few terms of \eqref{sesi} read
\begin{align}
S(\sigma)=&-\frac{i \left(\alpha ^2+g_1-g_2+1\right)}{\alpha -i} \sigma_1 
%
 \\
&-\tfrac{ \left(\alpha ^4-4 i \alpha ^3-2 \alpha ^2-4 i \alpha +g_1^2-2 \alpha ^2 g_1+4 i \alpha  g_1-2 g_1 g_2+2 g_1+g_2^2-2 \alpha ^2 g_2+4 i \alpha  g_2+2 g_2-3\right)}{8 (\alpha -i)^3 (\alpha -2 i)} \sigma_1^2\nn \\
&-\tfrac{ \left(3 i \alpha ^4+8 \alpha ^3-2 i \alpha ^2+8 \alpha -i g_1^2+6 i \alpha ^2 g_1+16 \alpha  g_1+2 i g_1 g_2-10 i g_1-i g_2^2-2 i \alpha ^2 g_2-8 \alpha  g_2+6 i g_2-5 i\right)}{4 (\alpha -i)^2 (\alpha -2 i)} \sigma_2 + \cdots . \nn
\end{align}


\providecommand{\href}[2]{#2}

\end{document}